\documentclass[aps,prd,amssymb,groupedaddress,showpacs]{revtex4}
\usepackage{graphicx,bm,color,psfrag,hyperref}
\usepackage{amsmath}
\usepackage{amssymb}
\usepackage{amsfonts}

\begin{document}

\title{\textbf{$f(R,{T_{\mu\nu} T^{\mu\nu}})$ gravity and Cardassian-like expansion as one of its consequences}}
\author{Nihan Kat{\i}rc{\i}}
\email{nihan.katirci@boun.edu.tr}

\author{Mehmet Kavuk}
\email{mehmet.kavuk@boun.edu.tr}

\affiliation{Department of Physics, Bo\u{g}azi\c{c}i University, Bebek Istanbul, Turkey}

\date{\today}

\begin{abstract}
We propose a new model of gravity where the Ricci scalar (R) in Einstein-Hilbert action is replaced by an arbitrary function of R and of the norm of energy-momentum tensor i.e., $f(R,T_{\mu\nu}T^{\mu\nu})$. Field equations are derived in the metric formalism. We find that the  equation of motion of massive test particles is non-geodesic and these test particles are acted upon by a force which is orthogonal to the four-velocity of the particles.
We also find the Newtonian limit of the model to calculate the extra acceleration
which can affect the perihelion of Mercury. There is a deviation from the general relativistic(GR)
result unless the energy density of fluid is constant. Arranging $\alpha$ parameter gives an opportunity to cure the inconsistency between the observational values for the abundance of light elements and the standard Big Bang Nucleosynthesis results. Even the dust dominated universe undergoes an accelerated expansion without using a cosmological constant in Model II. With this specific choice of $f(R,T_{\mu\nu}T^{\mu\nu})$, we get the a Cardassian-like expansion.

\end{abstract}

\pacs{04.50.Kd, 98.80.-k, 95.36.+x}
\maketitle

\section{Introduction}
\label{Intro}

Recent observations demonstrate that the universe expands
in an accelerated manner at the current epoch \cite{perlmutter,riess}. Even though the Einstein Field Equations (EFE),
$G_{\mu\nu}=8\pi GT_{\mu\nu},$
permit the expanding universe, it does not permit the universe undergoing accelerated
expansion without invoking some mysterious unknown component called dark energy.
Before explaining the possible candidates that make the universe expanding we briefly talk about the type of
modifications that can be made to EFE.
The generic action for our purpose is as follows:
$S=\int L_g \sqrt{-g}d^4x+\int L_m\sqrt{-g}d^4x$.
Here $L_g$ denotes the gravitational Lagrangian and $L_m$ stands for the matter Lagrangian. If we choose
gravitational Lagrangian as curvature scalar, $L_g=\frac{R}{16\pi G}$ our generic action becomes Einstein-Hilbert action and by
using the metric formalism we get EFE. To modify the EFE, we can either modify the
matter Lagrangian part of the generic action which shows itself as some additional sort of matter species on the
right hand side of EFE or we can modify the gravitational Lagrangian in our generic action in order to modify the left hand side of
EFE. As a matter of fact one can always write down EFE in the standard form
$G_{\mu\nu}=8\pi GT_{\mu\nu}$ by absorbing in $T_{\mu\nu}$ all the gravity modifications.

One possible candidate to explain the source of this accelerated expansion is called dark energy (DE). There have not been any convincing argument about
its origin. Dark energy differs from the other components of the universe such as baryonic matter and radiation,
in the sense that it has a negative pressure. It is this large negative pressure which creates a gravitational
repulsion to suppress the gravitational attraction and hence resulting in accelerated expansion. The actor of accelerated expansion
can be considered as the famous cosmological constant where constant refers to the energy density. By simply putting
this constant to the Einstein equations in a Lorentz invariant manner -which can be achieved by simply
multiplying it with the metric $( \Lambda g_{\mu\nu} )$ \cite{Zeldovich} gives rise to the desired result. Despite the success of
the cosmological constant as dark energy, it has a flaw which shows itself in explaining its value \cite{cosmoprob,sahni}. According to the observations, the energy density of cosmological constant must be of the order of
$\rho_{\rm{\Lambda}}\simeq 10^{-47}$ GeV$^4$. However from the perspective of a particle physicist, the origin of the
cosmological constant must be found in the vacuum energy density whose value is estimated to be $\rho_{\rm{vac}}\simeq
10^{74}$ GeV$^4$. This huge discrepancy between the values must be explained.

The late time cosmic acceleration can also be explained by $f(R)$ gravity theory where f is a function of Ricci
scalar $(R)$ \cite{RevNoOd}-\cite{Carroll:2003wy}. It corresponds to the choice $L_g=\frac{f(R)}{16\pi G}$ in our generic action. The conditions of the
viable cosmological models corresponds to f(R) gravity can be found in
\cite{vmcapozziellonojiri,vmnojiri,vmamarzguioui,vmkoivisto,vmstarobinsky,vmli,vmbergliaffa,vmsantos,vmcognola,vmfaraoni,
vmfaraoniNadeau,vmnojiriOdintsov,vmsokolowski,vmFaraoni,vmbohmer,vmcarloni,vmcapozziello,vmnojiriodintsovtretyakov,
vmNojiriodintsov,vmtsujikawa,vmananda}, and constraints obtained from the classical tests of GR for the Solar
System scale seem to rule out most of the models proposed so far
\cite{nojiri03,stchiba,sterickcek,stchibasmith,stnojiri,stcapozziello,Olmo07}. However some models passing Solar
System tests can be obtained
\cite{Hu:2007nk,stFaraoni,stfaulkner,stzhang,stpun,Sawicki:2007tf,Amendola:2007nt}.

Recently $f(R,T)$ gravity theory, where gravitational Lagrangian is given by an arbitrary function of the Ricci scalar $R$ and of the trace of the energy-momentum tensor $T$, received some attention \cite{fRT}. Actually this model can be seen as the application of more general theory where $L_g$ is given by an arbitrary function of Ricci scalar and of the matter Lagrangian, i.e., $f(R,L_m)$\cite{Bertolami:2007gv}. In these models the most remarkable part is that as matter is non-minimally coupled to the Ricci scalar, hence the motion of particles is non-geodesic and the particles are acted upon by a force which is orthogonal to their four-velocity.

 In this work we propose a new model of gravity where the gravitational Lagrangian of Einstein-Hilbert action $L_g=\frac{R}{16\pi G}$ where R is just a Ricci scalar is replaced by an arbitrary function of Ricci scalar of and the contraction of the energy-momentum tensor with itself i.e.,$f(R,T_{\mu\nu}T^{\mu\nu})$. We will follow almost the same line of reasoning as in \cite{fRT} and \cite{Bertolami:2007gv} to discuss our model's predictions.
In section II we derive the field equations of $f(R,{T_{\mu\nu} T^{\mu\nu}})$ gravity by using metric formalism and in part A we explicitly show that energy in general is not conserved in this gravity theory, it is either created or destroyed according to the sign of our model parameter $\alpha$. However in part B we show that by imposing the energy condition we get a regime where energy
conservation law emerges. In section III we choose two different specific functions $f(R,{T_{\mu\nu} T^{\mu\nu}})$ to study its cosmological implications and in section IV we use observational Helium and Deuterium abundances to put a constraint on the model parameter $\alpha$. In the last section we find the Newtonian limit for both models.

\section{The gravitational field equations of $f(R,T_{\mu\nu} T^{\mu\nu})$ gravity}
  The action for the $f(R,T_{\mu\nu} T^{\mu\nu})$ gravity is considered as
\begin{equation}
S=\int f(R,T_{\mu\nu}T^{\mu\nu})\sqrt{-g}d^4x+\int L_m\sqrt{-g}d^4x,
\end{equation}
where $g$ is the determinant of the metric, $R$ is the Ricci scalar, $T_{\mu\nu}$ is the energy-momentum tensor, and $L_m$ is matter Lagrangian density. Here $T_{\mu\nu}$ in the function $f(R,T_{\mu\nu}T^{\mu\nu})$ represents the same matter with the Lagrangian $L_m$. As an alternative $T_{\mu\nu}$ can be taken as a different matter which we do not consider here.
Varying the action with respect to the inverse metric we get
\begin{equation}\label{varaction}
\delta S=\int \left(f_R\delta R+f_{T^2}\delta( T_{\mu\nu}T^{\mu\nu})-\frac{1}{2}g_{\mu\nu}f\delta g^{\mu\nu}
+\frac{1}{\sqrt{-g}}\delta (\sqrt{-g}L_m)\right)\sqrt{-g}d^4x,\end{equation}
where
\begin{equation}
f_R(R,T_{\mu\nu}T^{\mu\nu})=\frac{\partial f}{\partial R}, \hspace{5mm}
f_{T^2}(R,T_{\mu\nu}T^{\mu\nu})=\frac{\partial f}{\partial(T_{\mu\nu}T^{\mu\nu})}.\end{equation}
The energy momentum tensor is defined as
\begin{equation}
T_{\mu\nu}=-\frac{2}{\sqrt{-g}}\frac{\delta(\sqrt{-g}L_m)}{\delta g^{\mu\nu}}
\end{equation}

If the Lagrangian density of matter solely depends on the metric components and not on their derivatives then we
have
\begin{equation}
T_{\mu\nu}=g_{\mu\nu}L_m-2\frac{\partial L_m}{\partial g^{\mu\nu}}.\end{equation}

Field equations derived from \eqref{varaction} is
\begin{equation}
f_RR_{\mu\nu}-\frac{1}{2}fg_{\mu\nu}+(g_{\mu\nu}\nabla_{\alpha}\nabla^{\alpha}-\nabla_{\mu}\nabla_{\nu})f_R=\frac{1}{2}T_{\mu\nu}-f_{T^2}\theta_{\mu\nu},
\label{genfieldeq}
\end{equation}
where
\begin{equation}
\theta_{\mu\nu}=\frac{\delta(T_{\alpha\beta}T^{\alpha\beta})}{\delta g^{\mu\nu}},
\end{equation}
\begin{equation}
\theta_{\mu\nu}=-2L_m\bigg(T_{\mu\nu}-\frac{1}{2}g_{\mu\nu}T\bigg)-TT_{\mu\nu}+2T_{\mu}^{\alpha}T_{\nu\alpha}-4T^{\alpha\beta}\frac{\partial^2
L_m}{\partial g^{\mu\nu}\partial g^{\alpha\beta}}.\label{thetamunu}\end{equation}

As is seen from the equation above $\theta_{\mu\nu}$ depends on matter Lagrangian explicitly. In this paper we
will only use the energy-momentum tensor of a perfect fluid
\begin{equation}\label{energymomentum}T_{\mu\nu}=(\rho+p)u_{\mu}u_{\nu}+pg_{\mu\nu},\end{equation} where $\rho$ is the energy density and $p$ is the thermodynamic pressure. As is known that the definition of matter Lagrangian giving the perfect fluid energy-momentum tensor \eqref{energymomentum} is not unique, for consistency $L_m=p$ is assumed, and so the second variation of the energy-momentum tensor \eqref{energymomentum} in \eqref{thetamunu} is null \cite{OdintsovGomez}. Thus we have

\begin{equation}\label{Newthetamunu}
\theta_{\mu\nu}=-2L_m\bigg(T_{\mu\nu}-\frac{1}{2}g_{\mu\nu}T\bigg)-TT_{\mu\nu}+2T_{\mu}^{\alpha}T_{\nu\alpha}.
\end{equation}

\subsection{Non-conservation of energy}
The first thing to note is that in this model the continuity equation
\begin{eqnarray}
\dot \rho+3H(\rho+p)\neq 0,
\end{eqnarray}
is not satisfied. The covariant divergence of \eqref{genfieldeq} is
\begin{align}\label{nonconservedenergy}
\nabla^{\mu}T_{\mu\nu}=-f_{T^2}g_{\mu\nu}\nabla^{\mu}(T_{\mu\nu}T^{\mu\nu})+2\nabla^{\mu}(f_{T^2}\theta_{\mu\nu}).
\end{align}
With the aim of obtaining the modified form of continuity equation, we contract the above equation with the four velocity $u^{\mu}$ of test particles and easily get
\begin{align}\label{nonconser}
\dot \rho+3H(\rho+p)=f_{T^2}\nabla_{0}(T_{\alpha\beta}T^{\alpha\beta})-2u^{\nu}\nabla^{\mu}(f_{T^2}\theta_{\mu\nu}).
\end{align}

As explicitly seen  from the above equation the RHS terms act as a source for the matter content of the universe and so the energy is not conserved.
However in the next section we see that for a specific choice of function $f(R,T_{\mu\nu} T^{\mu\nu}),$ there is a regime where total energy is conserved.
 Moreover, for different choices of $f(R,T_{\mu\nu} T^{\mu\nu}),$  we will investigate \eqref{nonconser} in the next section.
 \subsection{Conservation of total energy: Determination of special choice of function $f(R,T_{\mu\nu} T^{\mu\nu})$}
To obtain the conservative models, RHS of the \eqref{nonconservedenergy} must be equal to zero. As we need $\theta_{\mu\nu},$ we insert \eqref{energymomentum} in \eqref{Newthetamunu} and obtain
\begin{align}
\label{thetamunu}
\theta_{\mu\nu}=(-\rho^2-4\rho p-3p^2)u_{\mu}u_{\nu}.\end{align}
Following the argument given in \cite{alvarenga} we assume that the general function $f(R,T_{\mu\nu}T^{\mu\nu})$ has the form
\begin{align} f(R,T_{\mu\nu}T^{\mu\nu})=f_1(R)+f_2(T_{\mu\nu}T^{\mu\nu})\end{align}
and demand that the RHS of the \eqref{nonconservedenergy} to be equal to zero. Then we get
\begin{align} f_{2T^2}(3+11w^2+6w^4)+2T^2f_{2T^2T^2}(1+4w^2+3w^4)=0,\end{align}
where \begin{equation}
f_{2T^2}=\frac{df_{2}}{d(T_{\mu\nu}T^{\mu\nu})}\end{equation}
and $ f_{2}= f_2(T_{\mu\nu}T^{\mu\nu}).$
The general solution of this differential equation is
\begin{align}\label{conservedfunction} f_2(T_{\mu\nu}T^{\mu\nu})=c_1(T_{\mu\nu}T^{\mu\nu})^{\frac{1+3w^2}{-2(1+4w^2+3w^4)}}+c_2\end{align}
where $c_1$ and $c_2$ are integration constants.
For example, for equation of state $w=0$, i.e., dust, \eqref{conservedfunction} becomes
\begin{align}f_2(T_{\mu\nu}T^{\mu\nu})=c_1(T_{\mu\nu}T^{\mu\nu})^{-\frac{1}{2}}+c_2\end{align}
If we choose our function as $f(R,T_{\mu\nu}T^{\mu\nu})=\frac{1}{16\pi G}(R+\alpha(T_{\mu\nu}T^{\mu\nu})^{-\frac{1}{2}})$ then we have a conserved model. However the constant $\alpha$ has the dimension of $[M]^6$ which is not natural. Hence in the next section we will choose models where we need dimensionless constants.

\section{COSMOLOGICAL APPLICATIONS OF  $f(R,T_{\mu\nu} T^{\mu\nu})$ GRAVITY}
In this section, we will analyze some cosmological solutions of the theory by choosing appropriate function
$f(R,T_{\mu\nu} T^{\mu\nu}).$ Moreover throughout the paper we will assume that the universe is homogeneous and
isotropic with the matter content whose Lagrangian is given by $L_m=p$.
The geometry of space-time is described by the Friedmann-Lemaitre-Robertson-Walker (FLRW) metric, and for
flat space-like sections given by
\begin{equation}\label{FRWL}ds^2=-dt^2+a^2(t)(dx^2+dy^2+dz^2),\end{equation} where $a(t)$ is the scale factor of the
universe.
Now we will analyze different cases and their cosmological implications by fixing  the function $f(R,T_{\mu\nu}T^{\mu\nu})$

\begin{itemize}
\item Model I: $f(R,T_{\mu\nu}T^{\mu\nu})=\frac{R}{16\pi G}+\alpha\sqrt{T_{\mu\nu}T^{\mu\nu}}$
\end{itemize}
Here $G$ is the Newtonian gravitational constant and given as $G=\frac{1}{8\pi M^{2}_{Pl}}$ and $\alpha$ is a dimensionless  coupling parameter.
With this consideration, \eqref{genfieldeq} becomes
\begin{equation}\label{model1}
R_{\mu\nu}-\frac{1}{2}g_{\mu\nu}R=8\pi GT_{\mu\nu}+
8\pi G\alpha g_{\mu\nu}\sqrt{T_{\alpha\beta}T^{\alpha\beta}}
-8\pi G\alpha\frac{\theta_{\mu\nu}}{\sqrt{T_{\alpha\beta}T^{\alpha\beta}}},
\end{equation}
By using \eqref{thetamunu} and the metric given in \eqref{FRWL}, \eqref{model1} becomes
\begin{equation}
3H^2=8\pi G\rho~\bigg(1+\alpha\frac{4w}{\sqrt{1+3w^2}}\bigg)=\rho_{ \rm{eff}}\label{friedman1}
\end{equation}
\begin{equation}
2\dot H+3H^2=8\pi G\rho~\bigg(-w-\alpha\sqrt{1+3w^2}\bigg)=-p_{\rm{eff}}\label{friedman2}
\end{equation}
where $H=\frac{\dot a}{a}$ is the Hubble parameter and $w$ is the equation of state parameter (EoS) satisfying $p=w\rho.$ As $\alpha=0$ case corresponds to GR we will call the RHS of \eqref{friedman1}, \eqref{friedman2} as effective density and pressure respectively to distinguish the difference from the GR.

Inserting \eqref{friedman1}  into \eqref{friedman2}, we obtain
\begin{equation}
2\dot H+3H^2=C(\alpha,w)3H^2\label{friedman3},
\end{equation}
where
\begin{equation}
C(\alpha,w)=\frac{-w-\alpha\sqrt{1+3w^2}}{1+\alpha\frac{4w}{\sqrt{1+3w^2}}}.
\end{equation}
The modification vanishes at $\alpha=0$, and we consider the effective EoS is parameterized as $w_{\rm{eff}}=-C$ thus we relate $w_{\rm{eff}}$ and $w$ as,
\begin{equation}w_{\rm{eff}}=\frac{w+\alpha\sqrt{1+3w^2}}{1+\alpha\frac{4w}{\sqrt{1+3w^2}}}.\label{omegaeff}
\end{equation}
In order to obtain the Hubble parameter, \eqref{friedman3} is integrated
\begin{equation}
 H=\frac{2}{3(1+w_{\rm{eff}})t}.
 \label{Hdef}
 \end{equation}

Now if we use the energy-momentum tensor of perfect fluid given in \eqref{energymomentum} then \eqref{nonconser} becomes
\begin{align}\dot \rho +3H\gamma\rho=0,\end{align} where
\begin{align}\gamma=\frac{1+w+\alpha\frac{1+4w+3w^2}{\sqrt{1+3w^2}}}{1+\alpha\frac{4w}{\sqrt{1+3w^2}}}.\end{align}
The energy density varies depending on $\alpha$ as \begin{align}\rho=\rho_0(\frac{a}{a_0})^{-3\gamma}.\end{align}

\begin{figure}[h!]
\centering \includegraphics[width=9cm,height=7cm]{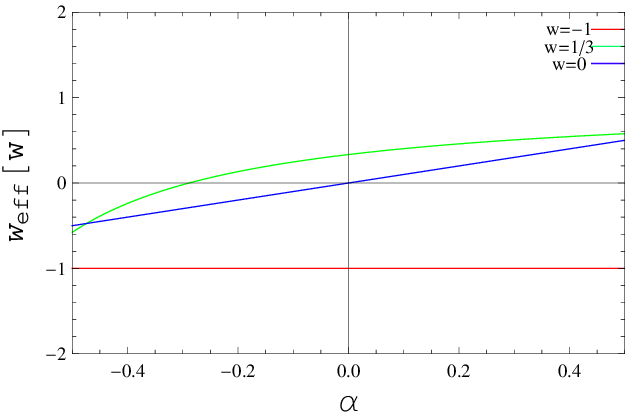}
\caption{The dependence of effective (EoS) on $\alpha$, the coupling of the
$\sqrt{T_{\mu\nu}T^{\mu\nu}}$ for different EoS in standard cosmology.}
\label{weffgraph}
\end{figure}

The dependence of the effective EoS in our proposed model on $\alpha$
for an interval $\alpha=[-0.5,0.5]$ in Figure \eqref{weffgraph}.

For example, for matter dominated universe $p=0$, thus $w=0$ ~~then $\gamma=1+\alpha$ giving $\rho_m=\rho_0a^{-3(1+\alpha)}$ which shows that the depending of the sign of $\alpha$ the matter in the universe will more slowly (if $\alpha$ is negative) or more quickly (if $\alpha$ is positive) dilute away than the expected evolution of energy density $\rho=\rho_0(\frac{a}{a_0})^{-3}.$

In the same manner $w=-1$,~~$\gamma=0$,~~$\rho_m=$const. meaning that the behavior of the density of dark energy doesn't change in this model and $w_{\rm{eff}}$ is independent of $\alpha$.

For radiation, we still have the same behavior as for matter since $w=\frac{1}{3}$~gives  $\gamma=\frac{\frac{4}{3}+\alpha\frac{4\sqrt{3}}{3}}{1+\alpha\frac{2\sqrt{3}}{3}}$~and
$\rho=\rho_0(\frac{a}{a_0})^{-4-\frac{4\alpha}{2\alpha+\sqrt{3}}}$ showing that the evolution of the density of radiation depends on the sign and the magnitude of $\alpha$.

In Section IV  we will put a constraint on $\alpha$ as it changes the behavior of energy densities from the expected energy behavior of $\Lambda$CDM.

 \begin{itemize}
\item Model II: $f(R,T_{\mu\nu}T^{\mu\nu})=\frac{1}{16\pi
    G}\bigg(R+\alpha(T_{\mu\nu}T^{\mu\nu})^{1/4}\bigg)$
\end{itemize}
First thing to be noted here is that the parameter $\alpha$ is still dimensionless as in Model I but G also couples
the additional term as a difference from that of Model I.
With this choice \eqref{genfieldeq} yields
 \begin{equation}\label{model2}
R_{\mu\nu}-\frac{1}{2}g_{\mu\nu}R=8\pi GT_{\mu\nu}+
\frac{\alpha}{2} g_{\mu\nu}(T_{\alpha\beta}T^{\alpha\beta})^{1/4}
-\frac{\alpha}{4}\frac{\theta_{\mu\nu}}{(T_{\alpha\beta}T^{\alpha\beta})^{3/4}}.
\end{equation}
For the FLRW metric flat space-like sections, Friedmann equations of motion becomes
\begin{equation}
3H^2=8\pi G\rho+\alpha\frac{\sqrt{\rho}}{2}\bigg(\frac{-1+4w-3w^2}{(1+3w^2)^{3/4}}\bigg),
\label{fried}
\end{equation}
\begin{equation}
2\dot H+3H^2=-8\pi Gw\rho-\alpha\frac{\sqrt{\rho}}{2}(1+3w^2)^{1/4}.
\label{fried2}\end{equation}

\eqref{fried} has the same form with
\begin{align}
\label{cardassianequation}
H^2=A\rho+B\rho^n,
\end{align} where $n=1/2$ in our case. \eqref{cardassianequation} is given in \cite{cardassian} in which authors explained the late time acceleration without the need of DE.
As is seen for early times as the second term on the right hand side of \eqref{cardassianequation} can be ignored, conventional cosmology can be obtained whereas  for late times the second term in \eqref{cardassianequation} dominates
giving the scale factor evolution as $a(t)=t^{\frac{2}{3n}}$  which gives an expansion for $n\leq2/3$.
It needs to be mentioned that for $\alpha(T_{\mu \nu} T^{\mu \nu})^\beta$ we still get the Cardassian term  for the general case and the relation $n=2\beta$ is obtained. Theory predicts that the parameter $\alpha$ has units of  $M^{2-8\beta}$ or $M^{2-4n}$ and it is easily realized that other choices except $\beta=\frac{1}{4}$ for model II will need dimensionful coefficients $\alpha$  in \eqref{model2} and $B$ in \eqref{cardassianequation} which makes the theory less natural. Considering the relation $n=2\beta$ and the accelerated expansion requirement ($n\leq2/3$), the condition  $\beta\leq\frac{1}{3}$ should be satisfied.
We show that $T_{\mu\nu}T^{\mu\nu}$ coupling naturally gives Cardassian-like accelerated expansion.

 We also note that in Cardassian-type expansion scenario they assumed that the continuity equation holds and the matter density evolves as $\rho=\rho_{0}(\frac{a}{a_0})^{-3}$ whereas in our case this situation is not valid. Still as will be seen, our model gives similar results although we do not use any hypothetical fluid in an ad hoc way. The deviation from standard GR is identified by the second term of \eqref{cardassianequation} and the best fit values for the Cardassian model parameters are given as $n=-1.33$, $z_{eq}=0.43$ and $\omega_M=0.076$ using supernova magnitude versus redshift measurements \cite{zhu03} which showed that the best fitted values for the two parameters ($n,z_{eq}$) of the Cardassian model gives lower matter density than the current value derived from the measurements of the cosmic microwave background anisotropy and galaxy clusters. Due to the violation in the conservation of the total energy, standard matter EoS parameter is not valid for our case and there arises a deviation from $w_m=0$.

When conservation of energy momentum tensor is imposed as done in Section IIB, $w_m=0$ can be safely used for the dust dominated case. For the matter dominated case, the conservation requires the special choice of function $f(T_{\mu\nu}T^{\mu\nu})=(T_{\mu\nu}T^{\mu\nu})^{-\frac{1}{2}}$. Using the relation $n=2\beta$ between the two parameters $n$ and $\beta$ which correlates the Cardassian model with this model, it is interesting that best fitted value $n=-1.33$ from Supernovae measurements is approximately close to $n=-1$ calculated from the $\nabla^{\mu}T_{\mu\nu}=0$ condition. Actually the fitting procedure should be followed for our proposal, the handicap is not to solve the system to find $H(z)$ function, and to use limiting behavior of the system instead. This is not the scope of this paper.
\eqref{fried} and \eqref{fried2} can be put into the form
\begin{equation}
\dot H(t)=K(w,\alpha)+L(w,\alpha)(1+H^2)+M(w,\alpha)\sqrt {1+H^2}.
\end{equation}
With the substitution $H=\sinh{u}$ where $u$ is just a parameter we obtain the following integral
\begin{equation}\label{model2solution}t=\int\frac{\cosh{u}~du}{K+M\cosh{u}+L\cosh^2{u}}+c.\end{equation}
\eqref{model2solution} can be solved but its solution is too complicated to analyze and interpret. Instead in the following we consider limit cases for our purpose.

It is evident that one may choose the terms such that the energy density is low and neglect the first terms on the rhs of \eqref{fried}-\eqref{fried2} compared with the second ones in the late time universe
\begin{equation}
8\pi G w\sqrt{\rho} \ll \frac{\alpha (1+3w^2)^{1/4}}{2},
\label{approx}
\end{equation}
where $\alpha$ and $w$ are order of one and the relation gives the  $\rho \ll M_p^2$ condition satisfied. Substituting this approximation, given in \eqref{approx}, on the solution of \eqref{fried}-\eqref{fried2}, we obtain the Hubble parameter as follows:
\begin{equation}
H(t)=\int \frac{-\sqrt{\rho}\alpha w}{(1+3w^2)^{3/4}}dt+c_1.
\label{appHubble}
\end{equation}
It is interesting remark that $w=0$ case gives the constant Hubble parameter that signs the de Sitter expansion phase.
In what follows, we make the analogy of \eqref{friedman3} and obtain the effective EoS parameter as:
\begin{equation}
w_{eff}=\frac{(1+3w^2)}{-3w^2+4w-1}.
\label{weffcase2}
\end{equation}
As explicitly seen, $w_{\rm{eff}}$ is independent of $\alpha$ as a basic difference from that of the model I, given in \eqref{omegaeff}.
In other words in this model even the dust dominated universe ($w=0$) undergoes an accelerated expansion ($w_{\rm{eff}}=-1$).
For the late time we don't need any hypothetical substance whose EoS parameter is $w=-1$ or cosmological constant.
Inflationary models can also be accommodated within extended
theories of gravity. One of the earliest $F(R)$ gravity attempt is $R^2$ Starobinsky inflation model \cite{staroref}
and its predictions are fully consistent with the Planck constraints \cite{PlanckXXII}.
Recently, the unified version of inflation with dark energy has also been proposed in the $F(R)$ gravity concept \cite{RevNoOd,nojiri03}. Here, in model II, an accelerating cosmological solution can be searched for the early time limit. Then the second term should be negligible when compared to the first in \eqref{fried}-\eqref{fried2} as $t \rightarrow 0$. One may check that this case corresponds to standard GR and $w=0$ case gives rise to $w_{eff}=0.$ Hence inflation is not obtained in this case. Likewise, in model I, \eqref{omegaeff} and \eqref{Hdef} demonstrate the deviation from standard GR. It can be deduced that the rapid early expansion is not included in Model I, too.

\section{Constraint from Big Bang Nucleosynthesis (BBN)}
We will now find how the abundances of light elements changes for Model I in terms of parameter
 $\alpha$ and put a constraint on it in order to check whether our model can be in accordance with the observations or not.
 To do this, we will use formulae given in \cite{GS,VSS} and follow the arguments of \cite{OT,Kahya}.
 \subsection{\textbf{$^4$He abundance}}

In GR, Friedmann equations (when $\alpha=0$ in \eqref{friedman1} and \eqref{friedman2}) yield the Hubble parameter as

 \begin{equation}H=\frac{2}{3(1+w_{\rm{rad}})t},\end{equation}
where $w_{\rm{rad}}=\frac{1}{3}$. On the other hand in our model this equation becomes
 \begin{equation}H=\frac{2}{3(1+w_{\rm{eff(rad)}})t}.
 \label{Heff}
 \end{equation}
 To find that how our model changes the abundances of primordial light elements during the Nucleosynthesis, we are
 interested in the ratio of our model's Hubble parameter to the Hubble parameter of GR during the early
 radiation dominated era. Putting \eqref{omegaeff} into \eqref{Heff}, S parameter is obtained as

\begin{equation}S = \frac{H_a}{H_{\rm{SBBN}}} = \frac{1+w_{\rm{rad}}}{1+\left(\frac{w_{\rm{rad}}+\alpha\sqrt{1+3w_{\rm{rad}}^2}}{1+\alpha\frac{4w_{\rm{rad}}}{\sqrt{1+3w_{\rm{rad}}^2}}}\right)}\label{S},
\end{equation}
where SBBN is abbreviation for the Standard Big Bang Nucleosynthesis.
The primordial abundances of the light elements (primordial D, $^{3}$He,
$^{4}$He, $^{7}$Li, T) depend on the baryon density and the expansion rate
of the universe \cite{GS,VSS}. The baryon density parameter is given by \cite{GS}

\begin{equation}\label{etaB}
\eta_{10} \equiv 10^{10} \eta_{B} \equiv 10^{10} \frac{n_{B}}{n_{\gamma}},
\end{equation}
where $\eta_{B}$ gives the baryon to photon ratio and we
can take $\eta_{10} \simeq 6$ \cite{eta10}.

Here we will use the expression for the $^{4}$He abundance given in \cite{GK, GS7}

\begin{equation}\label{Yp}
Y_{p} = 0.2485 \pm 0.0006 + 0.0016[(\eta_{10} - 6) + 100(S-1)].
\end{equation}

As is seen from the equation above, for $S=1$ we get the SBBN helium fraction which is

$Y^{SBBN}_{p} = 0.2485 \pm 0.0006.$

We will choose the parameter $\alpha$ of our model to fit the observed abundances of Helium,
\begin{align}
0.2561 \pm 0.0108=0.2485 \pm 0.0006 + 0.0016((\eta_{10} - 6) + 100(S-1)).
\label{helium}\end{align}

Since $\eta_{10} \simeq 6$, we find that the desired value of S must be $S=1.0475\pm0.1050$ in order the
Helium abundances to fit the observation. Moreover during the primordial Nucleosynthesis the universe was in radiation dominated era where the value of
equation of state parameter is $w=\frac{1}{3}$. From \eqref{omegaeff} we have
$w_{eff}=\frac{1}{3}+0.77\alpha+O(\alpha^2)$ and using this in \eqref{helium} we found the limit for $\alpha=-0.0785$. This model provides a great opportunity to solve the the problem about the difference between the observation (slightly greater than SBBN predictions) and SBBN predictions without proposing a new neutrino species or new physics (NP). Fine tuned $\alpha$ slightly decrease the Helium abundance and also gives a better deuterium abundance predictions without changing the standard system drastically.

\subsection{\textbf{Deuterium and Lithium-7 abundances}}

In order to calculate the Deuterium abundance we will use the expression based on a numerical best fit given in
\cite{GS}

\begin{equation}\label{yDp}
y_{Dp} = 2.60\left(1 \pm 0.06\right)\left(\frac{6}{\eta_{10}-6(S-1)}\right)^{1.6}.
\end{equation}
To find the SBBN value of $y_{Dp}$ we put $S=1$ and $\eta_{10}\simeq 6$ which gives the value of $y_{Dp}$ as
$y_{Dp}^{SBBN} = 2.60 \pm 0.16$.

By the same line of reasoning we have in the previous section we will find the value of S for which
 the Deuterium abundance fits the observation. From the Table I we see that the observed value of Deuterium abundance is $y_{Dp}=2.88 \pm 0.22$. Using \eqref{yDp} and by equating it with the observed value

\begin{align}2.88 \pm 0.22=2.60\left(1 \pm 0.06\right)\left(\frac{6}{\eta_{10}-6(S-1)}\right)^{1.6}\end{align}

we find $S=1.062\pm0.444$ and using \eqref{S} our parameter turns out to be $\alpha=-0.10109$.

Now we will fix the value of parameter of our model as $\alpha=-0.08$ and calculate the abundances by using this new value which can be seen in Table I.
\begin{table}[tp]
\begin{center}
\caption{The abundances He-4, Deuterium and Li-7 for different models}
\begin{tabular}{lccc}
Models and Data / Abundances:&$Y_{p}$&$y_{Dp}$&$y_{Lip}$ \\
\hline
Observational data:&\;\;\;\;\;$0.2561 \pm 0.0108$\cite{53}&\;\;\;\;\;$2.88 \pm 0.22$\cite{I}&\;\;\;\;\;$1.1 -
1.5$\cite{Asp} \\
                   &\;\;\;\;\; \\
SBBN model:\;\;\;\;\;\;\;&$0.2485 \pm 0.0006$&$2.60 \pm 0.16$&$4.82 \pm 0.48$ \\
                   &\;\;\;\;\; \\
$\rm{Model~I}:f(R,T_{\mu\nu}T^{\mu\nu})=R+\alpha\sqrt{T_{\mu\nu}T^{\mu\nu}}$ where $\alpha=-0.08$:\;\;\;\;\;\;\;&$0.2562 \pm 0.0083$&$2.81 \pm 0.17$& $4.59 \pm 0.44$ \\
                   &\;\;\;\;\; \\

\end{tabular}
\end{center}\label{t1}
\end{table}

For $y_{Lip}$, we will use the expression based on a numerical best fit given in \cite{GS} as
\begin{equation}
\label{yLip}
y_{Lip} = 4.82(1 \pm 0.10)(\frac{\eta_{10}-3(S-1)}{6})^{2}.
\end{equation}
 It can be clearly seen from the Table I, Lithium-7 abundance remains still a problem for this model. Although both SBBN and our predictions are far from the observations, $y_{Lip}$ is found slightly better fit to observations.

\textbf{\emph{\section{Equation of Motion of Test Particles for both Models I and II}}}

We will define projection tensor as $h_{\mu\nu}=u_{\mu}u_{\nu}+g_{\mu\nu}$ which is orthogonal to the
four-velocity of the test particles $h_{\mu\nu}u^{\mu}=0.$
We will use the energy-momentum tensor of the perfect fluid for our massive test particles given by
\eqref{energymomentum}, \begin{equation}\label{energymomentum1}T_{\mu\nu}=(\rho+p)u_{\mu}u_{\nu}+pg_{\mu\nu},\end{equation} where
$u^{\mu}$ is the four-velocity of the massive test particles satisfying $u_{\mu}u^{\mu}=-1.$
If we now take the covariant divergence of \eqref{energymomentum1} we get,

\begin{align}\nabla^{\mu}T_{\mu\nu}=\nabla^{\mu}(\rho+p)u_{\mu}u_{\nu}+(\rho+p)u_{\mu}\nabla^{\mu}u_{\nu}
+(\rho+p)u_{\nu}\nabla^{\mu}u_{\mu}+\nabla^{\mu}pg_{\mu\nu}\nonumber\end{align}

Multiplying the above equation with $h^{\nu\lambda}$ and using $h_{\mu\nu}u^{\mu}=0$ one finds
\begin{align}h^{\nu\lambda}\nabla^{\mu}T_{\mu\nu}=
(\rho+p)h^{\nu\lambda}u_{\mu}\nabla^{\mu}u_{\nu}+h^{\lambda}_{\mu}\nabla^{\mu}p.\nonumber\end{align}

Contracting the above equation with $g_{\lambda\alpha}$ we get,

\begin{align}g_{\lambda\alpha}h^{\nu\lambda}\nabla^{\mu}T_{\mu\nu}=
(\rho+p)u_{\mu}\nabla^{\mu}u_{\alpha}+\nabla^{\mu}ph_{\alpha\mu}\nonumber\end{align}

where we have used the condition $u_{\nu}\nabla^{\mu}u^{\nu}=0$ which can be obtained from the covariant divergence of $u_{\mu}u^{\mu}=-1.$ \\
\subsection{\textbf{Model I}}
If we take the covariant divergence of  \eqref{model1} and apply the same operations that we did above we get
\begin{align}g_{\lambda\alpha}h^{\nu\lambda}\nabla^{\mu}T_{\mu\nu}= -\alpha
h_{\alpha\mu}\nabla^{\mu}(\sqrt{\rho^2+3p^2})-\alpha\frac{\rho^2+4\rho
p+3p^2}{\sqrt{\rho^2+3p^2}}u_{\mu}\nabla^{\mu}u_{\alpha}.\nonumber\end{align}

Thus we have,

\begin{equation}\label{geodesic}u^{\mu}\nabla_{\mu}u^{\alpha}=-\frac{\alpha\nabla_{\mu}(\sqrt{\rho^2+3p^2})
+\nabla_{\mu}p}{\rho+p+\alpha\frac{(\rho+3p)(\rho+p)}{\sqrt{\rho^2+3p^2}}}h^{\mu\alpha}=f^{\alpha}\end{equation}

As can be seen from the above equation, there is an extra
force $f^\alpha$ acting on them which is orthogonal to their four-velocity $f^\alpha u_{\alpha}=0$ causing non-geodesic motion. When the parameter $\alpha$ of the model vanishes, this extra force reduces to the form of the standard general
relativistic fluid motion, i.e.,

\begin{align}f^\alpha=- \frac{\nabla_{\mu}p}{\rho+p}h^{\mu\alpha}.\nonumber\end{align}

\subsection{\textbf{Model II}}
Now if we take the covariant divergence of  \eqref{model2} and apply the same operations that we did for model I we get
\begin{equation}\label{geodesic2}u^{\mu}\nabla_{\mu}u^{\alpha}=-\frac{2\alpha\nabla_{\mu}{(\rho^2+3p^2)^{1/4}}
+32\pi G\nabla_{\mu}p}{32\pi G(\rho+p)+\alpha\frac{(\rho^2+4\rho p+3p^2)}{(\rho^2+3p^2)^{1/4}}}h^{\mu\alpha}=f^{\alpha}.
\end{equation}
Note that the resulting four-force is orthogonal to the four-velocity of the particles as in Model I and it again becomes
\begin{align}f^\alpha=- \frac{\nabla_{\mu}p}{\rho+p}h^{\mu\alpha}.\nonumber\end{align}
which is the standart general relativistic fluid motion when the parameter $\alpha$ goes to zero.

\subsection{Action for a Free Particle}
In Special or in General Relativity
dynamics of a free test particle can be determined by varying the action given by
\begin{align}\label{freeparticleaction}S=\int d\tau = \int\sqrt{-g_{\mu\nu}dx^{\mu}dx^{\nu}}.\end{align}
Variation of the above equation gives us the usual geodesic equation
\begin{align}u^{\alpha}\nabla_{\alpha}u^{\mu}=0.\end{align}
In what follows we will first make an assumption that the action governing the dynamics of test particles in a spacetime governed by our gravity model is

\begin{equation}\label{particleaction}
S=\int\sqrt{Q}\sqrt{g_{\mu\nu}u^{\mu}u^{\nu}}.
\end{equation}
Then we will show that the variation of \eqref{particleaction} gives us \eqref{geodesic} \cite{fRLm}.

To prove this result we start with the Lagrange equations corresponding to
the action given in \eqref{particleaction}
\begin{equation}
\frac{d}{ds}\left( \frac{\partial L_{p}}{\partial u^{\lambda }}\right) -%
\frac{\partial L_{p}}{\partial x^{\lambda }}=0.
\end{equation}

Since
\begin{equation}
\frac{\partial L_{p}}{\partial u^{\lambda }}=\sqrt{Q}u_{\lambda },
\end{equation}
and
\begin{equation}
 \frac{\partial L_{p}}{\partial x^{\lambda }}=\frac{1}{2} \sqrt{Q}g_{\mu \nu,\lambda }u^{\mu }u^{\nu }+\frac{
 1}{2} \frac{Q_{,\lambda }}{Q},
\end{equation}
a straightforward calculation gives the equations of motion of the particle as
\begin{equation}
\frac{d^{2}x^{\mu }}{ds^{2}}+\Gamma _{\nu \lambda }^{\mu }u^{\nu }u^{\lambda
}+\left( u^{\mu }u^{\nu }+g^{\mu \nu }\right) \nabla _{\nu }\ln \sqrt{Q}=0,
\end{equation}
and
\begin{align}\label{generalgeodesic}u^{\alpha}\nabla_{\alpha}u^{\mu}=-(u^{\mu }u^{\nu }+g^{\mu \nu })\nabla_{\nu }\ln
\sqrt{Q},\end{align}
which has the same form as of \eqref{geodesic}.
\subsubsection{\textbf{Model I}}
Identifying \eqref{generalgeodesic} with \eqref{geodesic} we have,

\begin{align}\label{1}\nabla_{\mu }\ln \sqrt{Q}=\frac{\alpha\nabla_{\mu}\sqrt{\rho^2+3p^2}
+\nabla_{\mu}p}{\rho+p+\alpha\frac{(\rho+3p)(\rho+p)}{\sqrt{\rho^2+3p^2}}}\end{align}

We will use the EoS of the form $p=w\rho$ for the pressure of the fluid where w satisfies the condition $w\ll1$. Therefore the conditions such as $\rho+p\sim\rho$ and $\rho^2+p^2\sim\rho^2$ are considered and these assumptions are used in RHS of \eqref{1},
\begin{equation}
\frac{\alpha\nabla_{\mu}\rho
+w\nabla_{\mu}\rho}{\rho+\alpha\rho}=\frac{\nabla_{\mu}\rho}{\rho}\bigg(\frac{\alpha+w}{1+\alpha}\bigg)=\nabla_{\mu}ln\rho^{\frac{\alpha+w}{1+\alpha}}\end{equation}
is obtained. Comparing with \eqref{1} we obtain

\begin{align}\sqrt{Q}= (c\rho)^{\frac{\alpha+w}{1+\alpha}},\end{align}

where c is an arbitrary constant of integration. The above equation can be written as $\sqrt{Q}=1+(\frac{\alpha+w}{1+\alpha})ln(c\rho)=1+U(\rho).$ Thus $U(\rho)$ is determined as

 \begin{align}\label{U}U(\rho)=(\frac{\alpha+w}{1+\alpha})ln(c\rho).\end{align}
We use this to find the extra term appearing in the Newtonian limit of the model. Starting from the usual line element in GR, we know that

\begin{align}ds=\sqrt{-g_{\mu\nu}dx^{\mu}dx^{\nu}}=\sqrt{-g_{00}dt^2-2g_{0i}dtdx^{i}-g_{ij}dx^{i}dx^{j}}.\end{align}
In the weak field approximation the metric for spherically symmetric static object has components
$g_{00}=-(1+2\phi)$ and $g_{0i}=0$. Thus the above equation becomes,

\begin{align}ds\sim (1+2\phi-v^2)^{1/2}dt\sim (1+\phi-v^2/2)dt\end{align}
For the action given in \eqref{particleaction}, using the weak field approximation corresponding equation becomes
\begin{equation}
S=\int\sqrt{Q}\sqrt{g_{\mu\nu}u^{\mu}u^{\nu}}\sim\int(1+U(\rho)+\phi-\frac{v^2}{2})dt
\end{equation}
whose variation gives us the equation of motion of the fluid to the first order approximation

\begin{align}\delta\int(1+U(\rho)+\phi-\frac{v^2}{2})dt=0.\end{align}

The total acceleration of the system, $\vec{a}$, is given as
\begin{align}\vec{a}=-\vec{\nabla}\phi-\vec{\nabla}U(\rho)=\vec{a}_N+\vec{a}_p+\vec{a}_E,\end{align}
where $\vec{a}_N=-\vec{\nabla}\phi$ is the Newtonian acceleration, $\vec{a}_p$ is the hydrodynamical acceleration and $\vec{a}_E$ is the supplementary acceleration induced by the matter-geometry coupling.

Using \eqref{U} we get
\begin{align}\vec{a}_p+\vec{a}_E=-\vec{\nabla}U(\rho)=-c\frac{\nabla p}{(1+\alpha)\rho}-c\frac{\alpha}{1+\alpha}\frac{\nabla \rho}{\rho},\end{align}

where the constant c can be chosen as $1+\alpha$ to have the hydrodynamical acceleration. With this choice, we get \begin{align}\vec{a}_E=-\alpha\frac{\vec{\nabla}\rho}{\rho},
\end{align}
which shows that if the energy density of the fluid is constant then the extra acceleration $\vec{a}_E$ is zero.
So our predictions will not differ from the standard GR and we do not put into constraint
on the model parameter $\alpha$.
\subsubsection{\textbf{Model II}}
Likewise if we identify \eqref{generalgeodesic} with \eqref{geodesic2} we get
\begin{align}\label{model2geodesic}\nabla_{\mu }\ln \sqrt{Q}=\frac{2\alpha\nabla_{\mu}{(\rho^2+3p^2)^{1/4}}
+32\pi G\nabla_{\mu}p}{32\pi G(\rho+p)+\alpha\frac{(\rho^2+4\rho p+3p^2)}{(\rho^2+3p^2)^{1/4}}}.\end{align}
With the same assumptions that we considered in the previous part, i.e., $w\ll1$, $\rho+p\sim\rho,$ and $\rho^2+p^2\sim\rho^2$,
\eqref{model2geodesic} becomes
\begin{equation}
\nabla_{\mu }\ln \sqrt{Q}\sim\frac{2\alpha\nabla_{\mu}\sqrt{\rho}
+32\pi Gw\nabla_{\mu}\rho}{32\pi G\rho+\alpha\rho^{3/2}}.\end{equation}
Following the same procedure as for Model I we get the function $U(\rho)$ as
\begin{align}U(\rho)=2(-1+w)\ln(\alpha+32\pi G\sqrt\rho)+\ln\rho.\end{align}
from which we obtain the extra acceleration
\begin{align}\label{extraacceleration}\vec{a}_E=-c(\rho_{0})\frac{\alpha}{\alpha\rho+32\pi G\rho^{3/2}}\vec{\nabla}\rho\end{align}
 where $c(\rho_{0})$ is a constant and $\rho_0$ is a fixed value of density around which we made the expansion to get the correct
 form of hydrodynamical accelaration $\vec{a_{p}}=-\frac{\vec{\nabla p}}{\rho}.$
We again see from \eqref{extraacceleration} that the extra acceleration is zero as the energy density of the fluid is constant.
Hence in the Newtonian limit the predictions of this model are not different than that of GR.

\section{Conclusions}

In this paper, we proposed a new gravity model where matter is non-minimally coupled to geometry via the contraction of the energy-momentum tensor with itself. We derived the field equations in metric formalism and for some cases we examined the cosmological implications. We had two considerations, in Model II, $G$ also couples the additional term as a difference from that of Model I and we concluded that only matter gives rise to the accelerated expansion without any need for the cosmological constant or hypothetical fluid. This case is similar to the Cardassian expansion model and we realized that the addition of the norm of energy-momentum tensor to the action automatically gives the expansion in an accelerated way. The condition that satisfies the energy conservation in our proposed model requires $\beta$ to be $-\frac{1}{2}$. This value seems to be consistent with the fitted values for the Cardassian model.
We checked whether model I can be in accordance with the observations or not. By using the observational values of primordial abundances of light elements during the BBN we have put a constraint on the parameter of Model I and by fixing it we recalculated the Helium and Deuterium abundances. This means that we are free to fine-tune $\alpha$ to Helium and Deuterium abundance without proposing New Physics (NP). We derived the equation of motion of massive test particles and showed that a force is acted upon them resulting in non-geodesic motion. By finding the Newtonian limit we showed that our models can not be tested by the precession of the perihelion of Mercury and there is no new limit on the model parameter $\alpha$.

\section{Acknowledgements}
This paper is dedicated to Professor Metin
Ar{\i}k on the occasion of his $65^{th}$ birthday. N. Kat{\i}rc{\i} thanks  Bo\u{g}azi\c{c}i University for the financial
support provided by the Scientific Research Fund with project no: $7128$. M. Kavuk also thanks  Bo\u{g}azi\c{c}i University
for the financial support provided by the Scientific Research Fund with project no: $6700$.

\end{document}